\documentclass[aps,floatfix,preprint,preprintnumbers,nofootinbib,superscriptaddress]{revtex4}
\usepackage{natbib}
\setcitestyle{authoryear,open={(},close={)}}

\usepackage[all]{xy}
\usepackage{amsmath,upgreek}
\usepackage{amssymb}

\usepackage{pdfpages}
\usepackage{color}
\usepackage{graphicx,epstopdf}
\usepackage[colorlinks=true,
linkcolor=red,
 urlcolor=black,
 citecolor=black,hyperindex]{hyperref}
\usepackage{lineno}
\usepackage[]{changes}
\usepackage[normalem]{ulem}

\makeatletter
\def\@fnsymbol#1{\ensuremath{\ifcase#1\or *\or \ddagger\or
		\mathsection\or \mathparagraph\or \|\or **\or \dagger\dagger
		\or \ddagger\ddagger \else\@ctrerr\fi}}
\makeatother

\begin{document}
%\linenumbers

\title{Setting up the physical principles of resilience in a model of the Earth System}

\renewcommand{\baselinestretch}{0}

\author{Orfeu Bertolami}
\email{orfeu.bertolami@fc.up.pt}
\affiliation{Departamento de F\'isica e Astronomia, Faculdade de Ci\^{e}ncias da
Universidade do Porto, Rua do Campo Alegre s/n, 4169-007, Porto, Portugal; Centro de F\'isica das Universidades do Minho e do Porto, Rua do Campo Alegre s/n, 4169-007, Porto, Portugal.}
\author{Magnus Nystr\"om}
\email{Magnus.Nystrom@su.se}
\affiliation{Stockholm Resilience Centre, University of Stockholm, Albanov\"agen 28, 106 91, Stockholm, Sweden.}

{\renewcommand{\baselinestretch}{1.2}

\vspace{2 cm}

\begin{abstract}

Resilience is a property of social, ecological, social-ecological and biophysical systems. It describes the capacity of a system to cope with, adapt to and innovate in response to a changing surrounding. Given the current climate change crisis, ensuring conditions for a sustainable future for the habitability on the planet is fundamentally dependent on Earth System (ES) resilience. It is thus particularly relevant to establish a model that captures and frames resilience of the ES, most particularly in physical terms that can be influenced by human policy\footnote{See page 4 for examples of strategies}. In this work we propose that resilience can serve as a theoretical foundation when unpacking and describing metastable states of equilibrium and energy dissipation in any dynamic description of the variables that characterise the ES. Since the impact of the human activities can be suitably gauged by the planetary boundaries (PBs) and the planet's temperature is the net result of the multiple PB variables, such as  $\text{CO}_2$ concentration and radiative forcing, atmospheric aerosol loading, atmospheric ozone depletion, etc, then resilience features arise once conditions to avoid an ES runaway to a state where the average temperature is much higher than the current one. Our model shows that this runaway can be prevented by the presence of metastable states and dynamic friction built out of the interaction among the PB variables once suitable conditions are satisfied. In this work these conditions are specified. As humanity moves away from Holocene conditions, we argue that resilience features arising from metastable states might be crucial for the ES to follow sustainable trajectories in the Anthropocene that prevent it run into a much hotter potential equilibrium state.

\end{abstract}


\date{\today}
\maketitle

\renewcommand{\baselinestretch}{1.2}

\section{Introduction}

Over the past decades the human imprint on the Earth System (ES) has been exceptional \citep{Steffen:2015a, Jouffray:2020}. While the mass of humans is only about $0.01\%$ of the total biomass, we have become a dominant force in shaping the face of Earth, including its atmosphere, biosphere, hydrosphere and lithosphere \citep{Vitousek:1997, Crutzen:2002,Ellis:2011,Foley:2011,Nystrom:2019}, and as of 2020 the global human-made mass surpasses the dry-weight of all living biomass \citep{Elhacham:2020}. Thus, humans have become a hyper-keystone species \citep{WormPaine:2016}, which rivals geological forces in influencing the trajectory of the ES \citep{Steffen:2018}

A major concern of these changes is the risk of crossing of so-called tipping-points, which refer to the critical threshold at which a small change or event triggers a significant and potentially irreversible (regime) shift in a system \citep{Lenton:2008}. Tipping-points have been observed in various systems, such as ecosystems (e.g. food webs,  benthic communities), social systems (e.g. norms, policy), economic systems (e.g. market-based economy) and technological systems (e.g. steam engine, smartphone, artificial intelligence) \citep{Scheffer:2001,Scheffer:2009,Nyborg:2016}. Over the past couple of decades there have been raising concerns around the existence of tipping-elements, which are large-scale components (subsystems) of the ES that may transgress a tipping-point \citep{Lenton:2008,Barnosky:2012}. Example of such tipping-elements include, the Greenland Ice Sheet, the Atlantic Meridional Overturning Circulation (AMOC), permafrost, monsoon systems, and the Amazon rainforest. Importantly, these tipping-elements interact, which may lead to a cascading behaviour of the entire ES \citep{Wunderling:2024}. The consequences of these dynamics for humanity could be colossal \citep{Steffen:2018}.

Clearly, knowledge about  tipping-points, where they are located, when they are approached and identifying ways to navigate away from them, are key challenges for humanity \citep{Barnosky:2012, Scheffer:2012}. Two broad frameworks that could help assist in this regard are planetary boundaries and resilience theory. The two are complementary in the sense that the planetary boundaries provide a quantitative assessment whereas the resilience framework adds a strong theoretical underpinning.

The planetary boundaries (PB) framework \citep{Rockstrom:2009,Steffen:2015,Richardson:2023} has been used to define global and regional limits in biophysical processes – ‘safe operating space’ – that must not be crossed if humanity is to stay away from systemic and potentially irreversible shifts in the ES. As such, the planetary boundaries framework serves as a "global dashboard," tracking humanity's collective impact on key environmental factors that threaten the Earth's ability to sustain human life. More recently, focus has been directed towards exploration of how different boundaries can interact and potentially cascade, thereby shrinking the safe operating space for future human impacts on the ES \citep{Lade:2020}. Importantly, while the PB framework highlights the presence of tipping points in biophysical processes, it does not specify their exact thresholds. Instead, it delineates two risk zones: a zone of increasing risk and a high-risk zone. In the former, the further boundary limits are exceeded, the greater the likelihood of causing significant harm—destabilizing critical Earth system processes and disrupting essential life-support functions. In the latter, or high-risk zone, there is a substantial risk of severe and potentially irreversible damage to key planetary functions. In essence, these zones are defined at a precautionary distance from the estimated locations of potential tipping points.

The resilience concept describes the extent to which a system can resist and develop (e.g. ecosystems or the the entire ES) with change by absorbing recurrent perturbations, deal with uncertainty and risk, and still sustain its key properties \citep{Holling:2001,Folke:2006}. This conception of resilience is based on the understanding that humans and Nature are deeply interconnected through feedbacks between social and ecological components, which together influence overall behavior and dynamics \citep{Biggs:2012}. This interdependence defines a social-ecological system \citep{Berkes:1998} in which human well-being and prosperity rely on the stability and functioning of the Earth system \citep{Folke:2011}. Multiple states (regimes), tipping-points and self-reinforcing feedback mechanisms (hysteresis) are a central feature of resilience \citep{Holling:2001}. In cases where resilience is high, a powerful shock – such as, storms, large wildfires, pest outbreaks in ecosystems, or armed conflicts, trade wars, and supply chain disruptions in social systems – is required to push the system beyond a tipping-point and into another state.	However, gradual (creeping) change – such as, loss biodiversity, habitat fragmentation and pesticide resistance in ecosystems, or growing inequality and changing social norms in society – erodes resilience of the current state. This makes the system vulnerable even to smaller perturbations. Once the system finds itself in this new state it can be difficult, or even impossible to reverse due to self-reinforcing feedback mechanisms \citep{Scheffer:2001,Scheffer:2009,Nystrom:2019}. Within the context of PB variables, species extinction (i.e. biodiversity loss PB) represents an irreversible process. Resilience has also been suggested as a conceptual framework that could assist in developing paths towards sustainability \citep{Folke:2016}. Hence, it can serve as a theoretical and practical foundation for the planetary boundaries framework. An important point to bear in mind, however is that resilience is a property of a system and is neither "good" nor "bad" per se. It can help maintain the current state of a system no matter whether it is deemed desirable or undesirable. The Holocene epoch has allowed development of agriculture, permanent settlements, and the emergence of complex human societies, so maintaining Holocene-like conditions can be deemed desirable, and safeguarding of resilience that support these conditions of critical importance for humanity \citep{Steffen:2018}.

Bearing in mind the resilience concept and its importance we aim in this work to specify, in the context of a thermodynamical model of the ES, what are the physical properties that manifest themselves collectively as resilience features of the ES. Our starting point is a thermodynamical model of the ES from Holocene state conditions to other potentially stable states, which can be regarded as phase transitions and admit a description through the Landau-Ginzburg Theory (LGT) \citep{Bertolami:2018,Bertolami:2019,Barbosa:2020}. The LGT is a theoretical framework used in physics to describe phase transitions, such as when a material changes from a solid to a liquid state or a magnetic material loses its magnetism. Here we use the LGT to describe the transitions the ES has gone throughout the history of  Earth. 

As we shall review in the next section, this framework allows for determining the equilibrium states of the ES in terms of the planet's biophysical subsystems or processes that are, due to the impact of the human activities, the driving forces that dominate its evolution. In the Anthropocene, human activities are  here collectively denoted by $H$. In the phase-transition model discussed in Refs. \citep{Bertolami:2018,Bertolami:2019,Barbosa:2020}, $H$ was considered an external field, however, in the present work, we admit that through policies and actions, the dynamic features of the ES can be altered so to modify the topographic landscape of possible Anthropocene trajectories. Way to do so include, mitigation strategies, such as halting deforestation and changing agricultural practices that contribute to $\text{CO}_2$ emission; transformation strategies, such as shifting from fossil fuel-based economies to ones based or renewable energy, and; restoration strategies, such as restoration of degraded ecosystems and $\text{CO}_2$ capture technologies.

As previously discussed, the proposed Landau-Ginzburg model allows for getting the evolution equation of the ES, the so-called Anthropocene equation, and to associate the sharp rise of  the physical parameters that characterise the ES to the great acceleration of the human activities \citep{Bertolami:2018}, which became  conspicuous from the second half of the 20th century and onwards \citep{Steffen:2015a}. 

However, as will be seen below, the original model did not exhibit explicit features that resemble resilience. This is the main purpose of the present work. As the model is based on thermodynamical arguments, one must seek for physical properties that would lead to a more resilient behaviour of the ES. In the context of the model, resilience is regarded as the resistance the ES shows in changing from one equilibrium state to another. At the present transient period, the Anthropocene, it has been hypothesised that the ES is moving away from the Holocene equilibrium state to a new state, potentially} a Hothouse Earth state \citep{Steffen:2018} (Fig. 1). As we shall see, our results show that resilience is associated to the existence of metastable states and explicit dissipation of energy that prevent the ES to runway towards the Hothouse Earth state.

%As we shall see, resilience can be associated to the existence of metastable states and retroactive mechanisms, which allow for the ES to settle in a stable/metastable state and show a considerable "resistance" to move away from this state.  

A pleasing feature of the proposed description is that it allows for drawing trajectories of the ES in the phase space of model's variables. By considering that the PBs and the ensued temperature display dynamics that are affected by PBs self-interactions which are shown to be different from zero \citep{Barbosa:2020}, two well defined and distinct sets of trajectories were identified upon assumptions about the evolution of the PB: a linear growth of the human activities, $H(T)=H_0 t$, where $H_0$ is an arbitrary constant, from which follows that all ES trajectories starting at the Holocene are led to Hothouse Earth state \citep{Steffen:2018} (Fig. 1) with a necessarily higher temperature than the Holocene average temperature \citep{Bertolami:2019}; if instead, the increase of the human activities impact on the ES obey a discrete logistic map \citep{May76,Jakobson81,Kingsland95}, trajectories can display bifurcations or chaotic behaviour \citep{Bernardini:2022}. Of course, as human activities are bounded by the finiteness of resources, the logistic map might be a more accurate description of its behaviour, although it is not quite clear what is the time span elapsed between successive steps of the logistic map. In any case, it is relevant to keep in mind that a too fast increase might give origin to trajectory bifurcations or even chaotic behaviour, which, of course, precludes predictions and control measures on the evolution of the ES. 
   
In this work we extend the previous studies of the ES model carried out in Refs. \citep{Bertolami:2018,Bertolami:2019,Barbosa:2020,Bernardini:2022} on various aspects. Previously, we aimed to show the inevitability of the Hothouse Earth state given the disestablishing nature of the human activities and the interplay among the PBs. Here, we consider the dynamic features arising from the self-interactions of the 9 identified PBs, here generically denoted as $h_i$,  $i= 1, ..., 9$, and show the specific conditions to implement resilience in the the eleven dimensional space $(\psi, h_i, F(\psi, h_i))$. Resilience can be regarded as a set of measures that prevent or delay the evolution of the ES towards a Hothouse Earth state and ensuring that this state is as close as possible to the Holocene state\footnote{Notice that prior the Anthropocene, the equilibrium states of the ES correspond to cooler (glaciation) and hotter (Hothouse Earth) equilibrium states with respect to the Holocene. However, at the Anthropocene, human activities lead inevitably the ES towards a Hothouse Earth state due to the massive emission of greenhouse gases. This materialises in the minus sign of the linear term in Eq. (\ref{eq:free_energy}) below.}. This can be implemented by creating metastable states to avoid a runaway situation due to a barrier that arises as higher-order terms into the Helmholtz free energy are introduced (cf. discussion below). A further requirement is dynamic friction, that is friction introduced via a kinetic energy-type term, to restrict the change of state in the phase space. This is a fairly natural condition as any realistic system dissipates energy. The specific conditions for the ES to acquire effective resilience features will be discussed below. Trajectories of the ES without and with resilience are depicted  in Figs. 1 and 2 respectively (cf. a detailed discussion below).    

This paper is organised as follows: in section II we review the cardinal aspects of the LGT of the ES and discuss the most relevant features of the dynamical system emerging from the model; in section III, we discuss the implementation of the resilience features in the model and connect them to properties that any model of the ES should have. Finally, in section IV we present our conclusions and discuss how our work can be extended to address several issues concerning features and transformation of the global social-ecological system.

%-----------------------------------------------------------------------------
%-----------------------------------------------------------------------------

\section{A Thermodynamical Model for the Earth System}

We first review the main features of the proposed model for the ES \citep{Bertolami:2018} and discuss in the next section the conditions to extend it in order to explicitly exhibit resilient properties.  

The proposal of Ref. \citep{Bertolami:2018} is to regard transitions of the ES as phase transitions which can be described by the LGT through an order parameter, $\psi$, and natural parameters (astronomical, geophysical, internal). In the Anthropocene, the natural forces average out to zero and the system is driven by the strength of the human activities, collectively denoted by $H$. In this approach, the  thermodynamic description of the system is obtained through the Helmholtz free energy, $F$, which can be written as an analytic function of an order parameter, $\psi$, which is chosen to be the reduced temperature relative to  Holocene average temperature, $\langle T_{\rm H}\rangle$, $\psi := (T - \langle T_{\rm H}\rangle)/ \langle T_{\rm H}\rangle$. Thus, in the Anthropocene, disregarding the spatial variation of $\psi$, one can write \citep{Bertolami:2018,Bertolami:2019}: 
\begin{equation}
F(\psi,H) = F_0 + a\psi^2 + b\psi^4 - \gamma H\psi, 
 \label{eq:free_energy}
\end{equation}
where $F_0$,  $a$, $b$ and $\gamma$ are constants. The linear term in $\psi$ corresponds to the human activities, which at the Anthropocene can match the quadratic and quartic contributions due to natural causes (astronomic, geological internal).  

The strength of the human activities are probed by their impact via the PBs \citep{Steffen:2015,Rockstrom:2009}, $h_i$, $i=1, 2, ..., 9$ with respect to their Holocene values. Given that the PB can interact among themselves, the most general expression for $H$ is given by \citep{Bertolami:2019}:
\begin{equation}
H = \sum_{i=1}^{9} h_i + \sum_{i,j=1}^{9} g_{ij} h_i h_j + \sum_{i,j,k=1}^{9} \alpha_{ijk} h_i h_j h_k +  \ldots,
\label{eqn:human_action}
\end{equation}
where $[g_{ij}]$ is a non-degenerate, $\det[g_{ij}] \neq 0$ $9 \times 9$ matrix. Similar conditions should be imposed on the coefficients $\alpha_{ijk}$ and $\beta_{ijkl} $ of the higher-order interaction terms. In principle, these interactions terms are sub-dominating, however, their importance has to be established empirically.  As pointed out in Ref.\,\citep{Bertolami:2019}, the interaction terms may lead to new equilibrium states and suggest some mitigation strategies depending on their sign and strength in the matrix entries  \citep{Bertolami:2019}. This will be explicitly discussed in the next section. In Ref. \citep{Barbosa:2020}, it was shown that the interaction term between the climate change variable, $\text{CO}_2$ concentration, say, $h_1$, and the oceans acidity, say, $h_2$, was non-vanishing and contributed to about $10\%$ of the value of the individual contributions themselves.

In order to introduce resilience features into the model, that is, resistance to change from one equilibrium state into another, we have to consider, contrary to previous works  \citep{Bertolami:2018,Bertolami:2019,Barbosa:2020,Bernardini:2022}, that the PBs are dynamical variables that are not only passively changed due to human activities, but that can be actively altered so to boost the resilience features of the ES. This allows us to project how the ES would behave depending on its initial state and subsequent trajectory in the phase space of the model, specified through the variables $(\psi,\dot{\psi}, h_i, \dot{h_i})$. Thus, for a given set of initial conditions, corresponding to a state $(\psi(0),\dot{\psi}(0), h_i(0),  \dot{h_i}(0))$ in the phase space, one can, in principle, obtain the trajectories, \emph{orbits}, in the phase space after solving the initial value problem through the evolution equations of the system. The equations of motion are obtained through the Lagrangian or equivalently through the Hamiltonian formalism. The latter, yielding to first order differential equations, is more suitable to establish a dynamical system in its canonical form. 

The Lagrangian function must include, besides the potential, which is given by the free energy, a set of kinetic energy terms for the canonical coordinates. The simplest possible kinetic term is a quadratic term proportional to the squared first derivative of each coordinate. Thus, we can write the following Lagrangian: 
{\setlength\arraycolsep{2pt}
\begin{eqnarray}
 \mathcal{L}(q,\dot{q}) = \frac{\mu}{2} \dot{\psi}^2 + \frac{\nu}{2} \sum_{i=1}^{9}\dot{h_i}^2 - F_0 - a\psi^2 - b\psi^4 + \gamma H\psi,
\end{eqnarray}}
where $\mu$ and $\nu$ are arbitrary constants and the dots stand for time derivatives. The constant $\nu$ is assumed to be the same for all PB variables.

Aiming to get the Hamiltonian function, we evince the relevant canonical conjugate momenta associated to $\psi$ and to a generic PB variable, $h_i$: 
\begin{equation}
 p_{\psi} = \frac{\partial \mathcal{L}}{\partial \dot{\psi}} = \mu \dot{\psi},
\end{equation}
\begin{equation}
 p_{h_i} = \frac{\partial \mathcal{L}}{\partial \dot{h_i}} = \nu \dot{h_i},
\end{equation}
from which follows the Hamiltonian function
\begin{equation}
 \mathcal{H}(\psi, p) = \frac{p_{\psi}^2}{2\mu} +  \sum_{i=1}^{9}\frac{p_{h_i}^2}{2\nu} + F_0 + a \psi^2 + b \psi^4 - \gamma H\psi,
\end{equation}
and Hamilton's equations,
\begin{equation}
 \dot{\psi} = \frac{\partial \mathcal{H}}{\partial p_{\psi}}, \quad \dot{p_{\psi}} = - \frac{\partial \mathcal{H}}{\partial \psi},
\end{equation}
\begin{equation}
 \dot{h_1} = \frac{\partial \mathcal{H}}{\partial p_{h_i}}, \quad \dot{p_{h_i}} = - \frac{\partial \mathcal{H}}{\partial h_i}.
\end{equation}
The equations of motion read, considering for while just the contribution from the lowest order terms in Eq. (\ref{eqn:human_action}): 
\begin{equation}
\mu \ddot{\psi} = - 2 a \psi - 4 b \psi^3 + \gamma H
\label{eq:motionpsi}
\end{equation}
and
\begin{equation}
\nu \ddot{h_i} =  \gamma \psi.
\label{eq:motionhi}
\end{equation}
To exemplify the behaviour of variables $\psi$ and $h_i$, let us obtain the resulting solutions for the simple case  considered in Ref. \citep{Bertolami:2019}. For $b \simeq 0$, we can neglect the cubic term in the equation of motion for $\psi$ to get the equation of an harmonic oscillator under the action of an external force, $H(t)$. This yields for the simple case of an initial linear time evolution,
\begin{equation}
 H(t) = H_0 t,
 \label{eq:linearH} 
\end{equation}
for an equilibrium initial state, $\dot{\psi}(0)=0$, the analytical solution:
\begin{equation}
 \psi(t) = \psi_0 \cos (\omega t) + \alpha t,
\label{eq:linearpsi} 
\end{equation}
where $\omega = \sqrt{2a / \mu} $ is an angular frequency, $\alpha = \gamma H_0 / 2 a$ and $\psi_0$ is an arbitrary constant fixed by the initial conditions.

The solution for the impact on the PB, $h_i(t)$, which initially behaves collectively as Eq. (\ref{eq:linearH}), that is $\sum_{i=1}^9 h_i (t \simeq 0) = H_0$, quickly evolves to a cubic growth in time: 
\begin{equation}
h_i(t) = A \cos (\omega t) + Bt^3 + \alpha_i t,
\label{eq:nlinearpsi} 
\end{equation}
where $A= -\gamma \psi_0/\nu \omega^2$, $B= \alpha \gamma/6 \nu$, for an arbitrary $\alpha_i$. 

These solutions show that if the temperature $\psi$ grows from an initial linear collective behaviour of the PBs, $H=H_0 t$, then it quickly drives the $h_i$s towards a cubic growth. Clearly, this model shows no resilience features as depicted in Fig. 1, where one clearly sees that from the Holocene, Anthropocene trajectories inevitably evolve towards a Hothouse Earth state.  

\begin{figure}
\includegraphics[width=1.0\columnwidth]{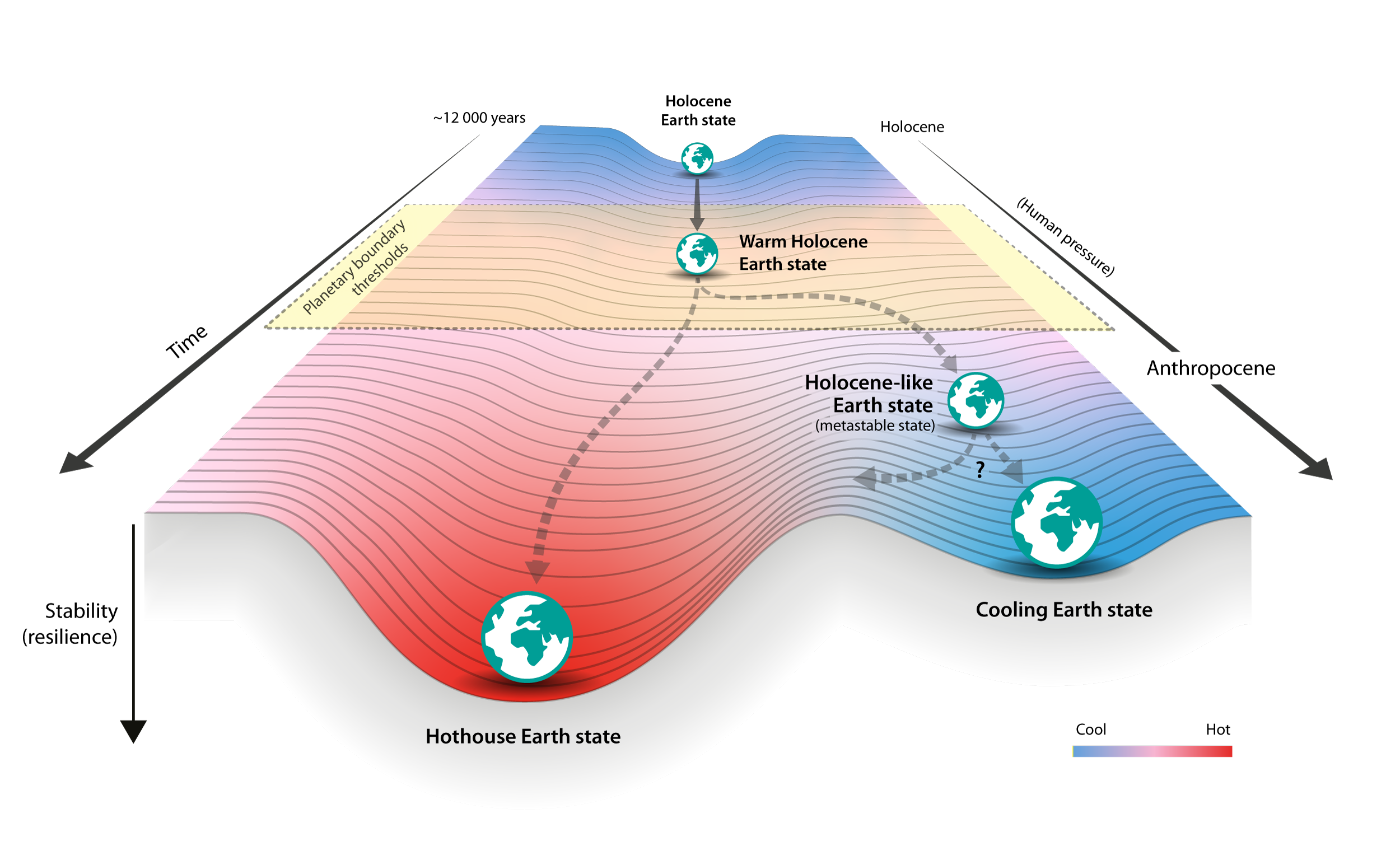}
\caption{A schematic illustration of the evolution of the Earth System with a start from the Neolithic revolution (~12.000 years ago). Leading up to its current state (i.e. "warm Holocene Earth state") 7 of 9 planetary boundaries have been transgressed. A continuation on this pathway suggests that the Earth system may end up in a Hothouse Earth state (Steffen et al. 2018) (left pathway). However, explicit dissipation of energy, and policies and actions geared at building resilience of a metastable "Holocene-like Earth state" (see also Fig. 2) could provide an opportunity to build a trajectory toward a future "cooling Earth state" (right pathway).}
\label{Noresilience}
\end{figure}

In what follows we shall consider the introduction into the free energy function of a cubic term for $\psi$ and higher than linear order terms for the PBs as these will allow for metastable states to arise, thus leading to bounded solutions for $\psi$ and the PBs. Metastable states correspond  to potential intermediate energy states between the Holocene state and the Hothouse Earth least energy state. In the LGT, metastable states can be considered and studied through cubic terms in the Helmholtz free energy. The conditions for the appearance of metastable states were already discussed in a completely different context, namely in a proposal to classify rocky planets \citep{Bertolami:2022}, using the ideias developed in Refs. \citep{Bertolami:2018,Bertolami:2019,Barbosa:2020} to describe the ES. In concrete terms, cubic terms might arise from PB interactions that have a strong dependence on the temperature. 

Before concluding this discussion it is worth stressing once again that the behaviour of the ES depends crucially on the assumptions about the evolution of the PB. Indeed, as pointed out in the introduction, the supposition that human activities grow linearly as in Eq. (\ref{eq:linearH}) implies, as exemplified above, that ES trajectories lead to a potential "Hothouse Earth" state \citep{Bertolami:2019} as discussed by Ref. \citep{Steffen:2018}. However, if the human activities impact on the ES behaves as a discrete logistic map \footnote{This means that the evolution of the PB, $h_i$, ($i=1,2,...,9$) is considered to be discrete and obey the equation $h_{i(j+1)}=r h_{i(j)}  \left( 1-\alpha h_{i(j)} \right)$, where $j$ denotes the number of "generations", $r$ is the rate of growth and $\alpha$ a constant.}, as suggested in Ref. \citep{Bernardini:2022}, then evolution will depend the rate of growth of human activities as solutions admit regular trajectories as well as trajectories that present bifurcations and even chaotic behaviour. In the next section we shall consider the features that must be introduced in the Helmholtz free energy and the conditions they must satisfy in order to avoid the ES evolves towards the Hothouse Earth state.

%-----------------------------------------------------------------------------
%-----------------------------------------------------------------------------

\section{Setting up the physical principles of resilience}
\label{resilience}

As mentioned above, in this model resilience features are associated to bounded trajectories in the Anthropocene and these ask for the existence of metastable states. In the LGT the metastable states arise by intruding cubic terms on the free energy. As pointed out in Ref. \citep{Bertolami:2022}, the introduction of a cubic term allows for a richer variety of equilibrium states. Indeed, consider the free energy: 
\begin{equation}
F(\psi,H) = F_0 + a\psi^2 - c |\psi|^3 + b\psi^4 - \gamma H\psi, 
 \label{eq:free_energy_cubic}
\end{equation}
where we assume that constants $b$, $c$ and $\gamma$ are positive, while constant $a$ can be negative.

The existence of extrema is given by two conditions. The first one reads: 
\begin{equation}
\frac{\partial F(\psi,H)}{\partial \psi} = 0 = 2a \psi - 3c \psi^2 + 4 b \psi^3 - \gamma H. 
 \label{eq:free_energy_cubic_minimum}
\end{equation}
The resulting cubic equation admits at least one real solution, say, $\psi_M$, meaning that there are at least two metastable states, $\psi_M$ and $-\psi_M$. Clearly, $\psi_M \not= 0$ as far as $H \not= 0$. 

However, the unboundedness of the evolution of the variables $(\psi, h_i)$ is due to the unboundedness of the PBs. Recent assessment of the PBs has shown that 7 out of the 9 PBs have gone beyond their Holocene values where they were at equilibrium, a state usually referred to as Safe Operating Space (SOS).

The motion in the eleven-dimensional configuration space, $(\psi, h_i, F(\psi, h_i))$, is quite complex, so in order to simplify the analysis we consider one single generic PB, $h_i$, and assume that the remaining ones are unchanged\footnote{Notice that the analysis of two-variables case is quite relevant as the Kolmogorov-Arnold representation theorem.establishes that any continuous function of several variables can be constructed out of a finite sum of two-variable functions.}. The free energy can be written explicitly in terms of the high order contributions in H depicted in Eq. (\ref{eqn:human_action}). We consider the essential set of terms in order to carry out the minimisation procedure, that is:
\begin{equation}
F(\psi,H) = \hat{F_0} + a\psi^2 - c |\psi|^3 + b\psi^4 - \gamma (h_i + g_{i} h_i^2 + b_{i} h_i^3)\psi, 
 \label{eq:free_energy_hcubic}
\end{equation}
where we have aggregated all contributions to the quadratic and cubic terms in $h_i$, {\bf a generic PB}, within the constants $g_{i}$ and $b_{i}$. To ensure boundedness it is necessary that $g_{i}$ is negative and that $b_{i}$ is positive. 

Thus, from Eq. (\ref{eq:free_energy_hcubic}), one gets the condition:
\begin{equation}
\frac{\partial F(\psi,h_i)}{\partial h_i} = 1 + 2  g_{i} h_i + 3  b_{i} h_i^2 = 0, 
 \label{eq:free_energy_cubic_hminimum}
\end{equation}
which admits real non-vanishing solutions, $h_{iM}$. as far as $g_i^2 > 3 b_1$ for $b_i \not= 0$ or $h_{iM} =- {1 \over  2 g_i}$ if $b_1= 0$. 

The general conditions to ensure that the extremum $(\psi_M, h_{iM})$ corresponds to a minimum and hence to a metastable state are given by: 
\begin{equation}
\frac{\partial^2F(\psi_M,h_{iM})}{\partial \psi^2}  \frac{\partial^2F(\psi_M,h_{iM})}{\partial h_i^2} -  \left(\frac{\partial^2F(\psi_M,h_{iM})}{\partial \psi \partial h_i}\right)^2 > 0. 
 \label{eq:free_energy_g1_minimum}
\end{equation}
and 
\begin{equation}
\frac{\partial^2F(\psi_M,h_{iM})}{\partial \psi^2}  > 0, 
 \label{eq:free_energy_g2_minimum}
\end{equation}   
which yield the following relationships:
\begin{equation}
g_i  <  - 3 b_i  h_{iM} 
 \label{eq:metastable1}
\end{equation}
and
\begin{equation}
2a -6 c |\psi_M| + 12 b \psi_M^2 > 0.
 \label{eq:metastable2}
\end{equation}
Satisfying these conditions imply the ES can settle in a the metastable state, $(\psi_M, h_{iM})$, that is, the system shows resilience and does not runaway towards the "Hothouse Earth" state as depicted in Fig. 2 as far as $3 b_i< g_i^2 < 9 b_i^2 h_{iM}$. .   

\begin{figure}
\includegraphics[width=1.0\columnwidth]{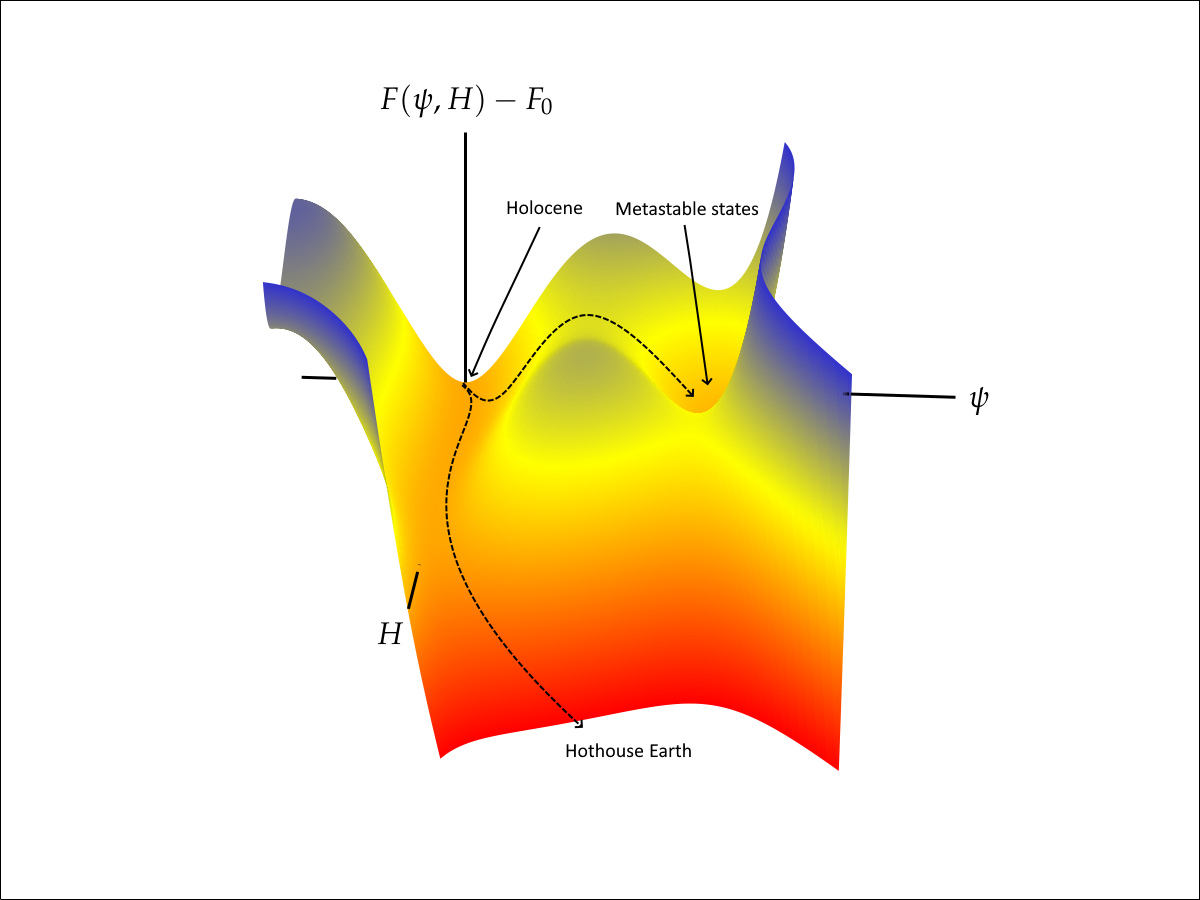}
\caption{Free energy in function of the temperature, planetary boundaries $(H)$ and resilience features (matastable state).}
\label{Resilience}
\end{figure}

Notice that the conditions for the existence of a metastable state can be met if $g_i < 0$ even if coefficients $b_{i}$ vanish. This is quite welcome as these coefficients are associated to higher-order interaction terms, which from phenomenological considerations, are presumably small. On the other hand, a non-vanishing and negative contribution from the quadratic term $h_i^2$ is absolutely necessary. Actually, the concrete case studied in Ref. \citep{Barbosa:2020} shows that this is possible. Furthermore, condition Eq. (\ref{eq:metastable2}) can be satisfied if $a < 0$.

Another feature associated to resilience is the ``inertia" that the ES shows in changing from a given state to another. This feature can be identified with the ubiquitous dissipation of energy present in any physical system. Most often dynamical dissipation processes can be described through velocity-proportional frictional forces which imply that just part of the free energy of a system is turned into kinetic energy, that is, motion of the system. In the Lagrangian/Hamiltonian formalism for a particle, the effect of these forces can be accounted through the Rayleigh dissipation function, $R = -\kappa p^2/m$, where $\kappa$ is a constant, $p$ is the canonical conjugate momentum and $m$ the mass of the particle. 

For the ES, introducing dissipation through the Rayleigh function implies that the left hand side of the equations of motion (\ref{eq:motionpsi}) and (\ref{eq:motionhi}) acquire extra terms $-\kappa_{\psi}  \dot{\psi}$ and 
$-\kappa_{h_i}  \dot{h_i}$, respectively. The effect of these terms is to reduce the amplitude of the motion of the ES once it goes from one state to another, thus acting as a resistance of the system to the change of its state. This can be clearly associated to resilience. 
 
These considerations are sufficient for setting the  physical conditions for  the resilience of the ES. As we have seen, a metastable state corresponding to the solution $(\psi_M, h_{iM})$ of 
equations (\ref{eq:free_energy_cubic_minimum}), (\ref{eq:free_energy_g1_minimum}), and (\ref{eq:free_energy_g2_minimum}), whose free energy (\ref{eq:free_energy_hcubic}) coefficients satisfy the conditions (\ref{eq:metastable1}) and (\ref{eq:metastable2}) together with the unavoidable dynamic friction /energy dissipation that exists in any system are the physical properties that endow the ES for having a resilient behaviour. For sure, further research is needed in order to establish which PBs are more suitable for setting up the conditions obtained above. This means that the PB properties concerning their dependence on the temperature and strength of their self-interaction and with other PBs must be further studied.

%Since the Holocene, the ES has been subjected to a great stress. From the Great Acceleration of the second half of the last century, which presumably sparked the Anthropocene, the hyper expansion of  human activities resulted that the safe operating space has been crossed for 6 of the 9 PBs \mbox{\citep{Richardson:2023}} and created all sorts of tensions, whose ongoing climate change crisis is the most persistent consequence for the ES. The tipping of some of the major ecosystems that compose the ES, such the Amazon rainforest and the Pacific Coral reefs, are already visible. As to the question of knowing if we have already inflicted an irreversible damage on ES or are close to it, only the understanding the mechanisms of resilience and how their boosting, through the PB interactions, can provide us with a knowledgeable answer. We hope that our work can provide a modest help in this respect.

%-----------------------------------------------------------------------------
%-----------------------------------------------------------------------------

\section{Conclusions}

In this work we have considered the physical principles to ascertain the conditions of resilience in a LGT model of the ES. In order to implement resilience features we have endowed and considered modifications of the free energy so to ensure the existence of metastable states. Furthermore, we have modelled the ES capability to remain in an equilibrium state by arguing that it can be suitably prevented to runway towards a potential Hothouse Earth state by the  presence of metastable states whose existence conditions were explicitly shown and the unavoidable dissipation of energy during the evolution of the relevant variables.  

Indeed, we have shown that, thanks to the PBs interactions, a metastable state $(\psi_M, h_{iM})$ can exist if the conditions, Eqs. (\ref{eq:metastable1}) and (\ref{eq:metastable2}), for the coefficients of the free energy, Eq. (\ref{eq:free_energy_hcubic}), are satisfied. As pointed out in the above discussion, these conditions can be satisfied even if coefficients $b_i$ vanish as far as $g_i<0$. 

Based on the observational data, it is possible to infer that the metastable state found above might correspond either to an actual state that the ES is close to reach or to a state that can be reached {by policy and actions (i.e. mitigation, transformation and restoration strategies) to drive the ES away from the Anthropocene traps it seems to be currently entangled in (see. Ref. \citep{Jorgensen} for a description of the 14 major Anthropocene traps). 

A recent assessment has shown that 7 out of the 9 PBs have been crossed \citep{Kitzmann2025} meaning that the evolution of most of the PBs is uncontrolled. Moreover it is unclear if the ES has already reached a point of no return, but it is evident that urgent measures to reverse the current development are needed. In fact, no single set of measures seems to be sufficient to halt the evolution of the PBs beyond the safe operating state. Two of the PBs that deserve particular attention are climate change and biosphere integrity. Both are deemed ``core" because their essential role in the ES. The climate system reflects the distribution and balance of energy at the Earth's surface, while the controls material and energy flows, helping to strengthen the systems's resilience against both rapid and long-term changes.   
This calls for a concerted action involving stewardship measures \citep{Steffen:2015a,Steffen:2011,OB:2022a}, bringing into the economy (internalising) the workings of the ES (see eg. Ref. \citep{OB:2024a}) and making them become part of revised economic paradigms \citep{Sureth:2023,OB:2023,OB:2025}, mitigation strategies that may include technological carbon sequestration (see e.g. \citep{OB:2024b,OB:2024c} and refs. therein),  and storage as means to curb climate overshoot, to avoid irreversible changes  to the ES that will compromise the navigation space for the future generations. 
Given that tipping in some major ecosystems constituting the Earth System is already being observed, such as in Pacific coral reef systems, and that other tipping elements including the Amazon rainforest and the polar ice sheets may be approaching the threshold of self-reinforcing tipping dynamics \citep{Lenton:2025}, a critical question arises as to whether irreversible damage to the Earth System has already occurred, or is imminent. The answer comes only through the understanding of the mechanisms of resilience and how their boosting, through the PB interactions, can be effective. We hope that our work can provide a modest help in this respect.

%Given that the tipping of some of the major ecosystems that compose the ES, such as the Amazon rainforest and the Pacific Coral reefs, are already visible, one faces the question of knowing if we have already inflicted an irreversible damage on ES, or are close to it. The answer comes only through the understanding of the mechanisms of resilience and how their boosting, through the PB interactions, can be effective. We hope that our work can provide a modest help in this respect.

\section*{Acknowledgements}

O.B. would like to thank the kind hospitality extended to him during his stay at the Stockholm Resilience Centre in May 2024 where the initial discussions that led to this work took place.  He is also grateful for the fruitful discussions on various matters with Lan Wang-Erlandsson, Peter Søgaard Jørgensen, Dieter Gerten, Ricardo El\'\i sio and Uno Svedin.

\section{Competing Interests}

The authors declare that they have no conflict of interest.

\section{Author Contributions}

OB and MN conceptualised the study. OB performed the formula calculations. OB and MN wrote and edited the paper.

\section{Financial Support}  

The work of Magnus Nystr\"om was supported by the Swedish Research Council grant (number 2020-04586).

%-- Bibliography -------------------------------------------------------------%

%{\color{red} Please, add as many references as you judge necessary. }

%\bibliographystyle{apalike}
\bibliography{RefOB}

@Article{Steffen:2015a,
	author = { Steffen, W. and Broadgate, W.  and Deutsch, L. and Gaffney, O. and Ludwig, C.},
	title = {The trajectory of the Anthropocene: The Great Acceleration.},
	journal = {Anthr. Rev.},
	year = {2015},
	number = {2},
	pages = {81–98}
}

@Article{Jouffray:2020,
  author  = {Jouffray, J and Blasiak, R and Norström, A and Österblom, H and Nyström, M},
  journal = {One Earth},
  title   = {The blue acceleration: the trajectory of human expansion into the ocean},
  year    = {2020},
  number  = {1},
  pages   = {43--54},
  volume  = {2},
  date    = {2020},
}

@Article{Vitousek:1997,
  author  = {Vitousek, P and Mooney, H and Lubchenco, J and Melillo, J},
  journal = {Science},
  title   = {Human domination of Earth’s ecosystems},
  year    = {1997},
  pages   = {494--499},
  volume  = {277},
  date    = {1997},
}

@Article{Crutzen:2002,
  author  = {Crutzen, P},
  journal = {Nature},
  title   = {Geology of mankind},
  year    = {2002},
  pages   = {23--23},
  volume  = {415},
  date    = {2002},
}

@Article{Ellis:2011,
  author  = {Ellis, E},
  journal = {Philos. Trans. R. Soc. A},
  title   = {Anthropogenic transformation of the terrestrial biosphere},
  year    = {2011},
  pages   = {1010--1035},
  volume  = {369},
  date    = {2011},
}

@Article{Foley:2011,
  author  = {Foley, J},
  journal = {Nature},
  title   = {Solutions for a cultivated planet},
  year    = {2011},
  pages   = {337--342},
  volume  = {478},
  date    = {2011},
}

@Article{Nystrom:2019,
  author  = {Nystr{\"o}m, M. and Jouffray, J. and Norstr{\"o}m, A. and Crona, B. and S{\o}gaard J{\o}rgensen, P. and Carpenter, S. and Bodin, {\"O} and Galaz, V. and Folke, C.},
  journal = {Nature},
  title   = {Anatomy and resilience of the global production ecosystem},
  year    = {2019},
  pages   = {98--108},
  volume  = {575},
  date    = {2019},
}

@Article{Elhacham:2020,
  author  = {Elhacham, E and Ben-Uri, L and Grozovski, J and Bar-On, Y and Milo, R},
  journal = {Nature},
  title   = {Global human-made mass exceeds all living biomass},
  year    = {2020},
  number  = {7838},
  pages   = {442--444},
  volume  = {588},
  date    = {2020},
}

@Article{WormPaine:2016,
  author  = {Worm, B and Paine, R},
  journal = {Trends in ecology and evolution},
  title   = {Humans as a hyperkeystone species},
  year    = {2016},
  pages   = {600--607},
  volume  = {31},
  date    = {2016},
}

@Article{Steffen:2018,
  author  = {Steffen, W and Rockstr{\"o}m, J and Richardson, K and Lenton, T and Folke, C and Liverman, D and Summerhayes, C and Barnosky, A and Cornell, S and Crucifix, M and Donges, J and Fetzer, I and Lade, S and Scheffer, M and Winkelmann, R and Schellnhuber, H},
  journal = {Proc Natl Acad Sci},
  title   = {Trajectories of the Earth System in the Anthropocene},
  year    = {2018},
  number  = {33},
  pages   = {8252--8259},
  volume  = {115},
  address = {USA},
  date    = {2018},
}

@Article{Scheffer:2001,
  author  = {Scheffer, M and Carpenter, S and Foley, J and Folke, C and Walker, B},
  journal = {Nature},
  title   = {Catastrophic shifts in ecosystems},
  year    = {2001},
  pages   = {591--596},
  volume  = {413},
  date    = {2001},
}

@Misc{Scheffer:2009,
  author    = {Scheffer, M},
  title     = {Critical transitions in nature and society},
  year      = {2009},
  date      = {2009},
  publisher = {Princeton University Press},
}

@Article{Nyborg:2016,
  author  = {Nyborg, K and Anderies, J and Dannenberg, A and Lindahl, T and Schill, C and Schl{\"u}ter, M and Adger, W and Arrow, K and Barrett, S and Carpenter, S and Chapin, Iii and F},
  journal = {Science},
  title   = {Social norms as solutions},
  year    = {2016},
  pages   = {42--43},
  volume  = {354},
  date    = {2016},
}

@InProceedings{Lenton:2008,
  author    = {Lenton, T and Held, H and Kriegler, E and Hall, J and Lucht, W and Rahmstorf, S and Schellnhuber, H},
  booktitle = {Proceedings of the National Academy of Sciences},
  title     = {Tipping elements in the Earth's climate system},
  year      = {2008},
  pages     = {1786--1793},
  volume    = {105},
  date      = {2008},
}

@Article{Barnosky:2012,
  author  = {Barnosky, A and Hadly, E and Bascompte, J and Berlow, E and Brown, J and Fortelius, M and Getz, W and Harte, J and Hastings, A and Marquet, P and Martinez, N},
  journal = {Nature},
  title   = {Approaching a state shift in Earth’s biosphere},
  year    = {2012},
  pages   = {52--58},
  volume  = {486},
  date    = {2012},
}

@Article{Wunderling:2024,
  author  = {Wunderling, N and Von Der Heydt, A and Aksenov, Y and Barker, S and Bastiaansen, R and Brovkin, V and Brunetti, M and Couplet, V and Kleinen, T and Lear, C and Lohmann, J},
  journal = {Earth System Dynamics},
  title   = {Climate tipping point interactions and cascades: a review},
  year    = {2024},
  number  = {1},
  pages   = {41--74},
  volume  = {15},
  date    = {2024},
}

@Article{Scheffer:2012,
  author  = {Scheffer, M and Carpenter, S and Lenton, T and Bascompte, J and Brock, W and Dakos, V and Van De Koppel, J and Van De Leemput, I and Levin, S and Van Nes, E and Pascual, M},
  journal = {Science},
  title   = {Anticipating critical transitions},
  year    = {2012},
  pages   = {344--348},
  volume  = {338},
  date    = {2012},
}

@article{Rockstrom:2009,
	author = {Rockstr{\"o}m, J. and Steffen, W. and Noone, K. and Persson, {\AA}. and Chapin III, F.S. and Lambin, E.F. and Lenton, T.M. and Scheffer, M. and Folke, C. and Schellnhuber, H.J. and Nykvist, B. and de Wit, C.A. and Hughes, T. and van der Leeuw, S. and Rodhe, H. and S{\"o}rlin, S. and Snyder, P.K. and Costanza, R. and Svedin, U. and Falkenmark, M. and Karlberg, L. and Corell, R.W. and Fabry, V.J. and Hansen, J. and Walker, B. and Liverman, D. and Richardson, K. and Crutzen, P. and Foley, J.A.},
	title = { A safe operating space for humanity}, 
	journal = {Nature},
	volume = {461},
	page = {472},
	year = {2009}
}

@Article{Steffen:2015,
  author  = {Steffen, W and Richardson, K and Rockstr{\"o}m, J and Cornell, S and Fetzer, I and Bennett, E and Biggs, R and Carpenter, S and De Vries, W and De Wit, C and Folke, C and Gerten, D and Heinke, J and Mace, G and Persson, L and Ramanathan, V and Reyers, B and Sorlin, S},
  journal = {Science},
  title   = {Planetary boundaries: Guiding human development on a changing planet},
  year    = {2015},
  number  = {6223},
  pages   = {1259855--1259855},
  volume  = {347},
  date    = {2015},
}

@Article{Richardson:2023,
  author  = {Richardson, K and Steffen, W and Lucht, W and Bendtsen, J and Cornell, S and Donges, J and Drüke, M and Fetzer, I and Bala, G and Von Bloh, W and Feulner, G and Fiedler, S and Gerten, D and Gleeson, T and Hofmann, M and Huiskamp, W and Kummu, M and Mohan, C and Nogués-Bravo, D and Petri, S and Porkka, M and Rahmstorf, S and Schaphoff, S and Thonicke, K and Tobian, A and Virkki, V and Weber, L and Rockstr{\"o}m, J},
  journal = {Science Advances},
  title   = {Earth beyond six of nine planetary boundaries},
  year    = {2023},
  pages   = {37},
  volume  = {9},
  date    = {2023},
}

@Article{Lade:2020,
  author  = {Lade, S and Steffen, W and De Vries, W and Carpenter, S and Donges, J and Gerten, D and Hoff, H and Newbold, T and Richardson, K and Rockström, J},
  journal = {Nature Sustainability},
  title   = {Human impacts on planetary boundaries amplified by Earth system interactions},
  year    = {2020},
  pages   = {119--128},
  volume  = {3},
  date    = {2020},
}

@Article{Holling:2001,
  author  = {Holling, C},
  journal = {Ecosystems},
  title   = {Understanding the complexity of economic, ecological, and social systems},
  year    = {2001},
  pages   = {390--405},
  volume  = {4},
  date    = {2001},
}

@Article{Folke:2006,
  author  = {Folke, C},
  journal = {Global Environmental Change},
  title   = {Resilience: The emergence of a perspective for social-ecological systems analyses},
  year    = {2006},
  pages   = {253--267},
  volume  = {16},
  date    = {2006},
}

@Article{Lenton:2025,
  author  = {Lenton, T and Milkoreit, M and Willcock, S and Abrams, F. and  Armstrong 
McKay, I and  Buxton, E. and  Donges, F. and Loriani, S and Wunderling, N and
Alkemade, F and Barrett, M and Constantino, S and Powell, T and Smith, R and
Boulton, A and  Pinho, P and Dijkstra, A and Pearce-Kelly, P and Roman-
Cuesta, M. and Dennis, D},
  journal = {},
  title   = {The Global Tipping Points Report, University of Exeter, UK.},
  year    = {2025},
  pages   = {},
  volume  = {},
  date    = {2025},
}

@Article{Folke:2016,
  author  = {Folke, C and Biggs, R and Norström, A and Reyers, B and Rockström, J},
  journal = {Ecology and Society},
  title   = {Social-ecological resilience and biosphere-based sustainability science},
  year    = {2016},
  number  = {3},
  volume  = {21},
  date    = {2016},
}

@Article{Bertolami:2018,
  author  = {Bertolami, O and Francisco, F},
  journal = {Global Planet. Change},
  title   = {A physical framework for the earth system, Anthropocene equation and the great acceleration},
  year    = {2018},
  pages   = {66--69},
  volume  = {169},
  date    = {2018},
}

@Article{Bertolami:2019,
  author  = {Bertolami, O and Francisco, F},
  journal = {Europhysics Letters},
  title   = {A phase-space description of the Earth System in the Anthropocene},
  year    = {2019},
  pages   = {59001},
  volume  = {127},
  date    = {2019},
}

@Article{Barbosa:2020,
  author  = {Barbosa, M and Bertolami, O and Francisco, F},
  journal = {The Anthropocene Review},
  title   = {Towards a Physically Motivated Planetary Accounting Framework},
  year    = {2020},
  number  = {3},
  volume  = {7},
  date    = {2020},
}

@Article{May76,
  author  = {May, R},
  journal = {Nature},
  title   = {Simple mathematical models with very complicated dynamics},
  year    = {1976},
  pages   = {459--467},
  volume  = {261},
  date    = {1976},
}

@Article{Jakobson81,
  author  = { Jakobson, M},
  journal = {Commun. Math. Phys},
  title   = {Absolutely continuous invariant measures for one-parameter families of one-dimensional maps},
  year    = {1981},
  pages   = {39--88},
  volume  = {81},
  date    = {1981},
}

@Misc{Kingsland95,
  author    = {Kingsland, S},
  title     = {Modeling Nature: Episodes in the History of Population Ecology},
  year      = {1995},
  address   = {Chicago},
  date      = {1995},
  publisher = {University of Chicago Press},
}

@Article{Bernardini:2022,
  author  = {Bernardini, A and Bertolami, O and Francisco, F},
  journal = {Evolving Earth},
  title   = {Chaotic Behaviour of the Earth System in the Anthropocene},
  year    = {2025},
  pages   = {100060},
  volume  = {3},
  date    = {2025},
}

@Article{Bertolami:2022,
  author  = {Bertolami, O and Francisco, F},
  journal = {Monthly Notices of the Royal Astronomical Society},
  title   = {Towards a classification scheme for the rocky planets based on equilibrium thermodynamic considerations},
  year    = {2022},
  pages   = {1037--1043},
  volume  = {515},
  date    = {2022},
}

@Article{Jorgensen,
  author  = { Søgaard Jørgensen, P and Jansen, R and Avila Ortega, D and Wang-Erlandsson, L and Donges, J and Österblom, H and Olsson, P and Nyström, M and Lade, S and Hahn, T and Folke, C and Peterson, G and Crépin, A-S},
  journal = {Philosophical Transactions B},
  title   = {Evolution of the polycrisis: Anthropocene traps that challenge global sustainability},
  year    = {2023},
  pages   = {1893},
  volume  = {379},
  date    = {2023},
}

@Article{Steffen:2011,
  author  = {Steffen, W and Persson, {aa} and Deutsch, L and Zalasiewicz, J and Williams, M and Richardson, K and Crumbly, C and Crutzen, P and Folke, C and Gordon, L and Molina, M and Ramanathan, V and Rockstr{"o}m, J and Scheffer, M and Schellnhuber, H and Svedin, U},
  journal = {Ambio},
  title   = {The Anthropocene: From Global Change to Planetary Stewardship},
  year    = {2011},
  number  = {7},
  pages   = {739--761},
  volume  = {40},
  date    = {2011},
}

@Article{OB:2022a,
  author  = {Orfeu Bertolami},
  journal = {Anthropocenica},
  title   = {Greening the Anthropocene},
  year    = {2022},
  volume  = {3},
  date    = {2022},
}

@Article{OB:2024a,
  author = {Orfeu Bertolami},
  title  = {Natural Capital as a Stock Option},
  journal = {arXiv:2404.14041},
  year   = {2024}
}

@Article{OB:2023,
  author  = {Bertolami, O and Gon\c{c}alves, C. D. },
  journal = {The Anthropocene Review},
  title   = {From a dynamic integrated climate economy (DICE) to a resilience integrated model of climate and economy (RIMCE)},
  year    = {2024},
  number  = {3},
  volume  = {11},
  date    = {2024},
}

@Article{OB:2025,
  author  = {Bertolami, O and Gon\c{c}alves, C. D. },
  journal = {The Anthropocene Review},
  title   = {Safety in an uncertain world within the Resilience Integrated Model of Climate and Economics (RIMCE)},
  year    = {2025},
  number  = {},
  volume  = {},
  date    = {2025},
}

@Article{Sureth:2023,
  author  = {Sureth, M and Kalkuhl, M and Edenhofer, O and Rockström, J},
  journal = {National {\"O}konomie und Statistik},
  title   = {A welfare economic approach to planetary boundaries. Jahrb{\"u}cher f{\"u}r},
  year    = {2023},
  number  = {5},
  pages   = {477--542},
  volume  = {243},
  date    = {2023},
}

@Article{Biggs:2012,
 author  = {Biggs, R. and Schl{\"u}ter, M. and Biggs, D. and Bohensky, E.L. and BurnSilver, S. and Cundill, G. and Dakos, V. and Daw, T.M. and Evans, L.S. and Kotschy, K. and et al},
 journal = {Annu. Rev. Environ. Resour.},
title   = {Toward Principles for Enhancing the Resilience of Ecosystem Services},
  year    = {2012},
  number  = {37},
  pages   = {421-448},
  volume  = {},
  date    = {2012},
}

@Article{Folke:2011,
  author  = {Folke, C. and Jansson, A. and Rockström, J and Olsson, P and Carpenter, S.R. and Chapin, F.S.  and  Crepin, A-S. and et al},
  journal = {Ambio},
  title   = {Reconnecting to the biosphere},
  year    = {2011},
  number  = {40},
  pages   = {719},
  volume  = {7},
  date    = {2011},
}

@InBook{OB:2024b,
	author = {Bertolami, O},
	title = {  Geoengineering and Climate Change: Methods, Risks, and Governance},
	chapter = {23 (Could the Well of an Orbital Lift be used to Dump Greenhouse Gases into Space?)},
	pages = {367--375},
	publisher = {John Wiley Sons},
	year = {2025}
}

@InBook{Berkes:1998,
	author = {Berkes, F and Folke, C },
	title = {Linking social and ecological systems: management practices and social mechanisms for building resilience},
	chapter = {},
	pages = {},
	publisher = {Cambridge University Press},
	year = {1998}
}

@Article{OB:2024c,
	author = {Bertolami, O. and de Matos, C.J.},
	title = {Cooling the Earth with \text{CO}2 filled containers in space},
	journal = {arXiv:/2401.07829.},
	year = {2024}
}

@article{Kitzmann2025,
	title = {Planetary Health Check 2025},
	year = {2025},
	author = {Sakschewski, B. and Caesar, L. and Andersen, L. S. and Bechthold, M. and Bergfeld L. and Beusen, A., L. and Billing, M. and Bodirsky, B. L. and Botsyun, S. and Dennis, D. P. and Donges, J. F. and Dou, X. and Eriksson, A. and Fetzer, I. and Gerten, D. and Häyhä, T. and Hebden, S. and Heckmann, T. and Heilemann, A. and Huiskamp, W.  and Jahnke, A. and Kaiser, J. and Krönke, J. and Kühnel, D. and Laureanti, N. C. and Li, C. and Liu, Z. and Loriani, S. and Ludescher, J. and Mathesius, S. and Norström, A. and Otto, F. and Paolucci, A. and Pokhotelov, D. and Rafiezadeh Shahi, K. and Raju, E. and Rostami, M. and Schaphoff, S. and Schmidt, C. and Steinert, N. J. and Stenzel, F. and Virkki, V. and Wendt-
Potthoff, K. and Wunderling, N. and Rockström, J.},
	publisher = {Potsdam Institute for Climate Impact Research (PIK)}
}

\end{document}